\documentclass[aps,preprint,epsfig]{revtex4}
\usepackage{graphicx,latexsym,amsmath}
\begin{document}
\title{Fluctuations of a long, semiflexible \\polymer in a narrow channel}
\author{Theodore W. Burkhardt,$^1$ Yingzi Yang,$^{2,\thinspace}$\footnote{Present address: Department of Macromolecular Science, Fudan University, Shanghai 200433, China} and Gerhard Gompper$^2$}
\affiliation{$^1$ Department of Physics, Temple University, Philadelphia, PA
19122, USA,$^2$ Institut f\"ur Festk\"orperforschung, Forschungszentrum
J\"ulich, D-52425 J\"ulich, Germany\\
}

\date{\today}
\begin{abstract}
We consider an inextensible, semiflexible polymer or worm-like chain, with persistence length $P$ and contour length $L$, fluctuating in a cylindrical channel of diameter $D$. In the
regime $D\ll P\ll L$, corresponding to a long, tightly confined polymer, the average length of the
channel $\langle R_\parallel\rangle$ occupied by the polymer and the mean square deviation from the average vary as $\langle R_\parallel\rangle=\left[1-\alpha_\circ(D/P)^{2/3}\right]L$ and $\langle\Delta R_\parallel^{\thinspace 2}\thinspace\rangle=\beta_\circ(D^2/P)L$, respectively, where
$\alpha_\circ$ and $\beta_\circ$ are dimensionless amplitudes. In earlier work
we determined $\alpha_\circ$ and the analogous amplitude $\alpha_\Box$ for
a channel with a rectangular cross section from simulations of very long chains. In this paper we estimate $\beta_\circ$ and $\beta_\Box$ from the simulations. The estimates are compared with exact analytical results for a semiflexible polymer confined in the transverse direction by a parabolic potential instead of a channel and with a recent experiment. For the parabolic confining potential we also obtain a simple analytic result for the distribution of $R_\parallel$ or radial distribution function, which is asymptotically exact for large $L$ and has the skewed shape seen experimentally.
\end{abstract}
\pacs{PACS}
\maketitle

\section{Introduction}
The statistical properties of biological polymers fluctuating in nano- or micro-channels have been studied in several recent experiments \cite{retal,ksp,kkp,kp,hlgr,chemsocrevpt,chemsocrevlc}. For biological polymers the persistence lengths are typically tens of nanometers or even larger. When the channel diameter is smaller than the persistence length, the stiffness of the polymer plays an important role. The polymer is stretched out in the channel with little backfolding, and the length of the channel occupied by the polymer is only slightly shorter than its contour length.

Measurements of the end-to-end distance of the polymer in a channel and its fluctuations provide information on the persistence and contour lengths of the polymer. This is of interest in studies of DNA fragments, for example, where sorting fragments of different length is desired, or in determining the change in bending rigidity upon binding of proteins \cite{sortingbinding1,sortingbinding2}.

In this paper we consider the simplest model for a confined biopolymer -  an inextensible, semiflexible filament or
worm-like chain with persistence length $P$ and contour length $L$ in a cylindrical
channel of diameter $D$. Here $D$ is an effective diameter, equal to twice the
maximum transverse displacement of the polymer from the symmetry axis of the
channel. For this system the distribution of the end-to-end distance or radial distribution function has been calculated theoretically \cite{lm,twf}, with the channel replaced by a parabolic confining potential, and studied with simulations \cite {cbb,twf}

 We will mainly consider the regime $D\ll P\ll L$, corresponding to a long, tightly confined polymer. In this regime the length of the channel $R_\parallel$ occupied by the polymer is essentially the same as the end-to-end distance. As discussed below, the distribution of $R_\parallel$ is Gaussian and is completely determined by the mean value $\langle R_\parallel\rangle$ and the mean square deviation $\langle \Delta R_\parallel^2\rangle$. These two quantities have simple scaling properties, summarized in the next paragraph. Our goal has been to determine the dimensionless proportionality constants in the scaling forms with good precision, so that one has unambiguous predictions for the worm-like chain that can be compared with experimental data and used, for example, to determine the persistence length.

In the regime $D\ll P\ll L$,
the free energy per unit length of confinement $\Delta f$, the average length of the
channel occupied by the polymer, and the variance or mean-square deviation from the average are given by
\begin{eqnarray}
&&\Delta f=A_\circ{k_BT\over P^{1/3}D^{2/3}}\;,\label{DeltaFcirc}\\
&&\langle R_\parallel\rangle=\left[1-\alpha_\circ\left({D\over P}\right)^{2/3}\right]L\;,
\label{Rparallelcirc}\\
&&\langle\Delta R_\parallel^{\thinspace 2}\thinspace\rangle=\beta_\circ{D^2\over P}\thinspace L\;,\label{DeltaRparallelcirc}
\end{eqnarray}
as follows from scaling arguments of Odijk \cite{to1,to2} and a detailed microscopic analysis \cite{ybg,twb97}.
For a channel with a rectangular cross section with edges $D_x$ and $D_y$,
\begin{eqnarray}
&&\Delta f=A_\Box{k_BT\over P^{1/3}}\left({1\over D_x^{2/3}}+
{1\over D_y^{2/3}}\right)\;,\label{DeltaFBox}\\
&&\langle R_\parallel\rangle=\left(1-\alpha_\Box{D_x^{2/3}+D_y^{2/3}\over P^{2/3}}\right)L\;,
\label{RparallelBox}\\
&&\langle\Delta R_\parallel^{\thinspace 2}\rangle=\beta_\Box{D_x^2+D_y^2\over P}\thinspace L\;.\label{DeltaRparallelBox}
\end{eqnarray}
Here $A_\circ$, $\alpha_\circ$, $\beta_\circ$, $A_\Box$, $\alpha_\Box$, and $\beta_\Box$ are dimensionless
universal amplitudes, which do not depend on $P$, $D$, $D_x$, and $D_y$.

The best estimates of the amplitudes in Eqs. (\ref{DeltaFcirc}), (\ref{Rparallelcirc}), (\ref{DeltaFBox}), and (\ref{RparallelBox}) to date are
\begin{eqnarray}
&&A_\Box=1.1036\;,\ 1.1038\pm 0.0006\;,\quad A_\circ=2.3565\pm
0.0004\;.\label{ABoxcirc}\\
&&\alpha_\Box=0.09137\pm
0.00007\;,\qquad\qquad\alpha_\circ=0.1701\pm 0.0001\;.\label{alphaBoxcirc}
\end{eqnarray}
The first entry for $A_\Box$ in Eq. (\ref{ABoxcirc}) was obtained by Burkhardt \cite{twb97}, by solving an integral
equation numerically, which arises in an exact analytical approach. The
other estimates are from our simulations \cite{ybg}
of very long polymers, with contour lengths up to $L\approx 1000\thinspace(2P)^{1/3}D^{2/3}$, where $\lambda\sim P^{1/3}D^{2/3}$ is the characteristic deflection length introduced by Odijk \cite{to1}.
Other estimates from simulations, compatible with these values but with larger
error bars, are given in Refs. \cite{bb,dfl,wg,cs}, and related results for a
helical polymer in a cylindrical channel in Ref. \cite{lbg}.

The paper is organized as follows. In Section II the underlying theoretical framework is reviewed, and new estimates from simulations,
\begin{equation}
\beta_\Box=0.0048\pm
0.0001\;,\qquad\qquad\beta_\circ=0.0075\pm 0.0002\;,\label{betaBoxcirc}
\end{equation}
for the amplitudes in Eqs. (\ref{DeltaRparallelcirc}) and Eqs. (\ref{DeltaRparallelBox}), obtained with same method as in Ref. \cite{ybg}, are presented.

In Section III and the Appendix we consider the mathematically more tractable problem of a polymer tightly confined in the transverse direction by a parabolic potential instead of a channel with hard walls. Exact analytic expressions for each of the quantities $\Delta f$, $R_\parallel$, and $\langle\Delta R_\parallel^{\thinspace 2}\rangle$ are derived. We find that $\langle\Delta R_\parallel^{\thinspace 2}\rangle$ is overestimated by about 30 $\%$ if the potential parameters are chosen to reproduce $L-\langle R_\parallel\rangle$ for a channel with hard walls. For the parabolic confining potential we also obtain a simple analytic result for the distribution of $R_\parallel$ or radial distribution function, which is asymptotically exact for large $L$ and for moderately large $L$ has the skewed shape seen experimentally.

In Section IV our predictions are compared with experimental results of K\"oster and Pfohl \cite{kp} for the radial distribution function of actin filaments in micro-channels. Section V contains closing remarks.

\section{Theoretical framework}
In the regime $D\ll P\ll L$, the line or filament by which we model the polymer is almost straight, without backfolding. Each such polymer configuration corresponds to a single valued
function $\vec{r}(t)$, where $(x,y,t)$ are Cartesian coordinates, and
$\vec{r}=(x,y)$ specifies the transverse displacement of the polymer from the
symmetry axis or $t$ axis of the channel. Since the slope $\vec{v}=d\vec{r}/dt$ with respect to the $t$ axis satisfies
$|\vec{v}|\ll 1$, the relation $L=\int_0^{R_\parallel}dt\left[1+\vec{v}(t)^{\;2}\right]^{1/2}$ between the contour length $L$ and the longitudinal length $R_\parallel$ may be replaced by
\begin{equation}
R_\parallel=L-{1\over 2}\int_0^L dt\;\vec{v}(t)^{\;2}\;,\label{extension}
\end{equation}
and the Hamiltonian ${\cal H}$ of the worm-like chain {\cite{explain} simplifies to
\begin{equation}
{{\cal H}\over k_BT}=\int_0^L dt\left[{P\over 2}\left({d^2\vec{r}\over dt^2}\right)^2+V(\vec{r})\right]\;.\label{hamiltonian}
\end{equation}
Here the two terms in square brackets are the bending energy per unit length and the confining potential per unit length, both divided by $k_BT$. For a polymer in a channel with hard walls, $V(\vec{r})$ takes the values $0$ and $\infty$ for $\vec{r}$ inside and outside the channel, respectively.

According to Eq. (\ref{extension}),  the average length of tube occupied by the polymer
and its variance or mean square deviation are given by
\begin{eqnarray}
&&\langle R_\parallel\rangle =L-{1\over 2}\int_0^L dt\;\langle\vec{v}(t)^{\;2}\rangle\;,\label{Rparallelav}\\
&&\langle \Delta R_\parallel^{\thinspace 2}\thinspace\rangle ={1\over 4}\int_0^L dt_1\int_0^L dt_2\;\left[\langle\vec{v}(t_1)^{\;2}\vec{v}(t_2)^{\;2}\rangle-
\langle\vec{v}(t_1)^{\;2}\rangle\langle\vec{v}(t_2)^{\;2}\rangle\right]\;,\label{meansquare}
\end{eqnarray}
where $\Delta R_\parallel=R_\parallel-\langle R_\parallel\rangle$.

For a tightly confined polymer in a channel with a rectangular cross section, the displacements of the polymer in the $x$ and $y$ directions are statistically independent. The partition function $Z$ factors into a product of two partition functions $Z_xZ_y$, which only involve displacements in the $x$ and $y$ directions, respectively. This is a consequence of the additive property $\left(d^2\vec{r}/dt^2\right)^2=\left(d^2x/dt^2\right)^2+\left(d^2y/dt^2\right)^2$ in the Hamiltonian (\ref{hamiltonian}) and the rectangular boundary, which does not break the statistical independence in the two transverse directions. From this and from rescaling lengths according to $x'=D_x^{-1}\thinspace x$, $t'=(2P)^{-1/3}D_x^{-2/3}\thinspace t$, it follows that the statistical averages on the right-hand sides of Eqs. (\ref{Rparallelav}) and (\ref{meansquare}) can all be determined from simulations of a long polymer with persistence length $P'={1\over 2}$ confined to the two dimensional strip $0<x'<1$ in the $(x',t')$ plane, as carried out in Ref. \cite{ybg}.

The statistical averages in Eqs. (\ref{Rparallelav}) and (\ref{meansquare}) can be expressed in terms of the variable
\begin{equation}
\zeta={1\over t'}\int_0^{t'}dt'\thinspace v_x'(t')^2\;,\label{zeta}
\end{equation}
where $v_x'^2=(dx'/dt')^2$.  According to the scaling transformations in the preceding paragraph,
\begin{equation}
\langle\zeta\rangle=
\left({2P\over D_x}\right)^{2/3}{1\over L}\int_0^L dt\thinspace\langle v_x(t)^{\;2}\rangle
=\left({2P\over D_y}\right)^{2/3}{1\over L}\int_0^L dt\thinspace\langle v_y(t)^{\;2}\rangle\;.\label {v'sqav}
\end{equation}

As discussed in the final paragraph of the Appendix, the quantity $\zeta$ defined in Eq. (\ref{zeta}) is expected to follow a Gaussian distribution for sufficiently large $L$, with the mean value
in Eq. (\ref{v'sqav}) and with variance $w^2$ given by
\begin{equation}
w^2=\langle\left(\zeta-\langle\zeta\rangle\right)^2\rangle={1\over t'^2}\int_0^{t'}dt_1'\int_0^{t'}dt_2'\thinspace\left[\langle v_x'(t_1')^2\thinspace v_x'(t_2')^2\rangle-\langle v_x'(t_1')^2\rangle\thinspace\langle v_x'(t_2')^2\rangle\right]\;.\label{w}
\end{equation}
The distribution determined from our simulations of polymers with values of $t'$ up to 300, shown in Fig. 1, is indeed very nearly Gaussian, and the variance $w^2$, as shown in Fig. 2, is in excellent agreement with the scaling behavior $w^2t'\to k$ for large $t'$, where $k$ is a constant, expected \cite{explainwidth} from Eq. (\ref{DeltaRparallelBox}). Substituting this relation in Eq. (\ref{w}) and expressing the scaled lengths in terms of the original variables gives
\begin{eqnarray}
k&=&{2P\over D_x^2}\thinspace{1\over L}\int_0^L dt_1\int_0^L dt_2\thinspace\left[\langle v_x(t_1)^{\;2}v_x(t_2)^{\;2}\rangle-\langle v_x(t_1)^{\;2}\rangle\langle v_x(t_2)^{\;2}\rangle\right]\nonumber\\
&=&{2P\over D_y^2}\thinspace{1\over L}\int_0^L dt_1\int_0^L dt_2\thinspace\left[\langle v_y(t_1)^{\;2}v_y(t_2)^{\;2}\rangle-\langle v_y(t_1)^{\;2}\rangle\langle v_y(t_2)^{\;2}\rangle\right]\;.\label{kdef}
\end{eqnarray}

According to our earlier paper \cite{ybg}, $\langle\zeta\rangle=0.2901\pm 0.0003$, and from the data shown in Fig. 2 of this paper, we estimate $k=0.0382\pm 0.0010$. Inserting these values in the relations $\alpha_\Box=2^{-5/3}\langle\zeta\rangle$ and $\beta_\Box={1\over 8}\thinspace k$, which follow from Eqs. (\ref{RparallelBox}), (\ref{DeltaRparallelBox}), and (\ref{Rparallelav})-(\ref{kdef}), we obtain the predictions for $\alpha_\Box$ and $\beta_\Box$ in Eqs. (\ref{alphaBoxcirc}) and (\ref{betaBoxcirc}).

The entries for $\alpha_\circ$ and $\beta_\circ$ in Eqs. (\ref{alphaBoxcirc}) and (\ref{betaBoxcirc}) follow, in a very similar way, from the result $\langle\zeta_\circ\rangle=0.5400\pm 0.0004$, where
\begin{equation}
\zeta_\circ={1\over t'}\int_0^{t'}dt'\thinspace\vec{v'}(t')^2\;,\label{zetacirc}
\end{equation}
obtained in Ref. \cite{ybg} from simulations of a polymer with longitudinal length $t'$ and persistence length $P'={1\over 2}$ in a channel with a circular cross section of diameter $D'=1$, and from the corresponding estimate $k_\circ=0.0602\pm 0.0020$, where $w_\circ^2=\langle\left(\zeta_\circ-\langle\zeta_\circ\rangle\right)^2\rangle\to k_\circ t'^{-1}$ for large $t'$.

\section{Polymer confined by parabolic potential}
Next we consider a polymer tightly confined in the transverse direction by a parabolic potential of the form
\begin{equation}
V(\vec{r})={1\over 2}\left(b_x\thinspace x^2+b_y\thinspace y^2\right)\label{parabolicpotential}
\end{equation}
instead of a channel with hard walls. The partition function $Z(\vec{r},\vec{v};\vec{r_0},\vec{v_0};t)$ corresponding to the Hamiltonian (\ref{hamiltonian}) with the parabolic potential energy (\ref{parabolicpotential}) was evaluated for arbitrary values of the position and slope, $(\vec{r},\vec{v})$ and $(\vec{r_0},\vec{v_0})$, at the polymer endpoints and arbitrary longitudinal length $t$ in Ref. \cite{twb95}.

The case $b_x=b_y$ of equal potential parameters has been studied by Levi and Mecke \cite{lm} and Th\"uroff {\it et al.} \cite{twf}, who calculated the distribution of $R_\parallel$ or radial distribution function and compared their predictions with the experiments of Ref. \cite{ksp,kp}. In this paper we consider distinct values of $b_x$ and $b_y$, as is appropriate for rectangular channels with $D_x\neq D_y$, and concentrate mainly on the large-$L$ limit and on the prediction of the 6 dimensionless amplitudes $A_\circ$, $\dots$, $\beta_\Box$ in Eqs. (\ref{DeltaFcirc})-(\ref{DeltaRparallelBox}).

Since the thermal averages in Eqs. (\ref{Rparallelav}) and (\ref{meansquare}) are integrated over the entire length of the polymer, the particular boundary conditions at the endpoints of the  polymer are unimportant in the large-$L$ limit.  Straightforward calculations, given in the Appendix, lead to the results
\begin{eqnarray}
&&\left(k_BT\right)^{-1}\Delta f=2^{-1/2}P^{-1/4}\left(b_x^{1/4}+b_y^{1/4}\right)\;,\label{Deltafparabolic}\\
&&\langle R_\parallel\rangle =\left[1-2^{-5/2}P^{-3/4}\left(b_x^{-1/4}+b_y^{-1/4}\right)\right]L\;,\label{Rparallelavparabolic}\\
&&\langle \Delta R_\parallel^{\thinspace 2}\thinspace\rangle =2^{-9/2}P^{-5/4}\left(b_x^{-3/4}+b_y^{-3/4}\right)L\;.\label{meansquareparabolic}
\end{eqnarray}

To obtain an approximate formula for the amplitude $\beta_\Box$ for a channels with hard walls and a rectangular cross section, defined in Eq. (\ref{DeltaRparallelBox}), we choose the parabolic potential parameters $b_x$ and $b_y$ in Eq. (\ref{Rparallelavparabolic}) so that the average longitudinal length $\langle R_\parallel\rangle$ in the channel, given by Eq. (\ref{RparallelBox}), is reproduced, term by term. Substituting these potential parameters in Eq. (\ref{meansquareparabolic}) and comparing with Eq. (\ref{DeltaRparallelBox}) leads to a formula for $\beta_\Box$ in terms of $\alpha_\Box$. This calculation and a similar one for the channel with a circular cross section lead to the relations
\begin{equation}
\beta_\Box=8\alpha_\Box^3\ ,\quad \beta_\circ=2\alpha_\circ^3\;.\label{parabolicapprox}
\end{equation}
We note that Eq. (\ref{parabolicapprox}) also follows from choosing the parabolic potential parameters in Eq. (\ref{meansquareparabolic}) to reproduce $\langle \Delta R_\parallel^{\thinspace 2}\rangle$ in Eqs. (\ref{DeltaRparallelcirc}) or (\ref{DeltaRparallelBox}), substituting these potential parameters in Eq. (\ref{Rparallelavparabolic}), and comparing the result with Eqs. (\ref{Rparallelcirc}) or
(\ref{RparallelBox}).

Substituting the values of $\alpha_\Box$ and $\alpha_\circ$ in Eq. (\ref{alphaBoxcirc}) into Eq. (\ref{parabolicapprox}), we obtain the predictions $\beta_\Box=0.00610\pm 0.00002$ and $\beta_\circ=0.00984\pm 0.00002$, which are 27 $\%$ and 31 $\%$ larger, respectively, than our estimates (\ref{betaBoxcirc}) from simulations.
Thus, we see that calculations in which the hard wall potential of is replaced by a softer, parabolic confining potential tend to overestimate the endpoint fluctuations $\langle \Delta R_\parallel^{\thinspace 2}\rangle$ if the potential parameters are chosen to reproduce $L-\langle R_\parallel\rangle$ for a channel with hard walls. Similarly, if the potential parameters are chosen to reproduce $\langle \Delta R_\parallel^{\thinspace 2}\rangle$ for a channel with hard walls, the quantity $L-\langle R_\parallel\rangle$ is underestimated.

The asymptotic forms of {\em both} $L-\langle R_\parallel\rangle$ and $\langle \Delta R_\parallel^{\thinspace 2}\rangle$ for a polymer in a channel with hard walls are correctly reproduced if not only the potential parameters, but also the persistence length $\tilde{P}$ of the equivalent parabolically confined polymer is properly chosen. Setting Eqs. (\ref{Rparallelavparabolic}) and (\ref{meansquareparabolic}), with $\tilde{P}$ in place of $P$, equal to the corresponding expressions (\ref{Rparallelcirc}), (\ref{DeltaRparallelcirc}), (\ref{RparallelBox}), and (\ref{DeltaRparallelBox}), and solving for $\tilde{P}$, we obtain
\begin{equation}
{\tilde P}={\beta_\Box\over 8\alpha_\Box^3}\thinspace P ,\quad\tilde{P}={\beta_\circ\over 2\alpha_\circ^3}
\thinspace P\label{Ptilde}
\end{equation}
for the rectangular and circular channel cross sections, respectively, where the same combinations of exponents occur as in Eq. (\ref{parabolicapprox}). Substituting the values of $\alpha_\Box$, $\alpha_\circ$, $\beta_\Box$, and $\beta_\circ$ from Eqs. (\ref{alphaBoxcirc}) and (\ref{betaBoxcirc}) in Eq. (\ref{Ptilde}), we find that the persistence length $\tilde{P}$ of the equivalent parabolically confined polymer is 21 $\%$ and 24 $\%$ smaller than the persistence length $P$ of the polymer in the rectangular and circular channel, respectively.

Finally, in the Appendix we derive simple analytic results,
in terms of ``inverse Gaussian" functions, for the radial distribution function of a polymer confined by a parabolic potential in the moderate to large $L$ regime. The predictions, given in Eqs. (\ref{invgauss})-(\ref{variance}), (\ref{PBox(xi)}), and (\ref{Pcirc(xi)}), with $\xi=L-R_\parallel$, are compared with experimental results for polymers in channels in the next section.

\section{Comparison with experiment}
Experiments on unconfined filaments of the biopolymer actin (see Ref. \cite{ksp} and references therein) have yielded estimates of 8 to 25 $\mu$m for the persistence length. With fluorescence microscopy K\"oster, Pfohl, and coworkers  \cite{ksp,kp} have measured the radial distribution of actin filaments with contour length $L= 21$ $\mu$m in channels with rectangular cross sections with depth $D_x=1.4$ $\mu$m and widths $D_y=1.5$, 4.0, 5.8, and 9.8 $\mu$m. Comparing their experimental results for the radial distribution function, shown below in Figs. 3 and 4, with the theoretical prediction of Levi and Mecke \cite{lm} for a parabolic confining potential, K\"oster, and Pfohl \cite{kp} find good agreement, for all four channels, with the value $P= 13\;\mu$m.

Since $L$ is only moderately larger than $P$, the above experimental parameters do not clearly satisfy $D_x,D_y\ll P\ll L$, the condition under which our predictions for $\langle R_\parallel\rangle$ and $\langle\Delta R_\parallel^{\thinspace 2}\rangle$ apply. Nevertheless it is interesting to compare the experiments with our predictions for the scaling regime.

As discussed above and in the last paragraph of the Appendix, the distribution of $R_\parallel\thinspace$ is expected to be Gaussian in the scaling regime, with mean value and variance given by Eqs. (\ref{RparallelBox}), (\ref{DeltaRparallelBox}), (\ref{alphaBoxcirc}), and
(\ref{betaBoxcirc}). Using these relations and the above experimental values of $L$, $D_x$, and $D_y$ to determine the mean and variance as a function of $P$, we have carried out least square fits of the experimental results to Gaussian distributions for all four channels, varying $P$ to optimize the fits. This leads to the results shown in Fig. 3, and the estimates $P = 7.61$, 11.1, 14.1, and 10.1 $\mu$m for the the channels with widths $D_y=1.5$, 4.0, 5.8 and 9.8 $\mu$m. The first two of these estimates are expected to be the most reliable, since the condition $D_x,D_y\ll P\ll L$ is more nearly satisfied.

We have also carried out fits of the experimental results in which both $\langle R_\parallel\rangle$ and $\langle \Delta R_\parallel^{\thinspace 2}\rangle$ are treated as fit parameters. In the large-$L$ limit these quantities yield two independent predictions,
\begin{eqnarray}
&&P=\left(\alpha_\Box\thinspace{D_x^{2/3}+D_y^{2/3}\over 1-\langle R_\parallel\rangle/L}\right)^{3/2}\;,\label{P1}\\
&&P=\beta_\Box\thinspace{D_x^2+D_y^2\over\langle\Delta R_\parallel^{\thinspace 2}\rangle/L}\;,\label{P2}
\end{eqnarray}
for the persistence length, which follow from solving Eqs. (\ref{RparallelBox}) and
(\ref{DeltaRparallelBox}) for $P$.

Least square fits of the same experimental data to the inverse Gaussian distribution, given by Eqs. (\ref{invgauss}) and (\ref{xi3d}), with both the mean $\langle R_\parallel\rangle$ and variance $\langle\Delta R_\parallel^{\thinspace 2}\rangle$ adjusted to optimize the fit, are shown in Fig. 4. Of course, the two-parameter fit reproduces the experimental distribution more closely than the one-parameter fit in Fig. 3. Both the inverse Gaussian distribution and a convolution of inverse Gaussian functions, as described in the Appendix, have the skewed form seen in the experimental data and lead to nearly the same results.

The fits shown in Fig. 4 lead to the estimates $P=(7.0,\thinspace 2.8)$, $(9.5,\thinspace 3.6)$, $(10.8,\thinspace 4.1)$, and $(7.2,\thinspace 3.2)$ in $\mu$m for the channel widths $D_y=1.5$, 4.0, 5.8, and 9.8 $\mu$m, where the first and second numbers in parenthesis follow from substituting the mean and variance from the best fit in Eqs. (\ref{P1}) and (\ref{P2}), respectively, with $\alpha_\Box$ and $\beta_\Box$ given by Eqs. (\ref{alphaBoxcirc}) and (\ref{betaBoxcirc}). All of these estimates are smaller than the values $P=13$ $\mu$m and $P=15\pm 3$ $\mu$m proposed in Refs. \cite{kp,lm}, respectively, and for each channel the estimate based on Eq. (\ref{P2}) is only 3 or 4 $\mu$m, less than half of the corresponding estimates based on Eq. (\ref{P1}). Determining the mean and variance by fitting the experimental data to an ordinary Gaussian distribution instead of an inverse Gaussian distribution or by evaluating the mean and variance directly from the experimental histograms without assuming a particular distribution leads to quite similar estimates.

Finite-size corrections probably account, at least in part, for the discrepancy in the estimates of $P$ based on Eqs. (\ref{P1}) and (\ref{P2}), with the smaller estimate coming from Eq. (\ref{P2}). As the contour length $L$ increases and the polymer is tightly confined over a greater fraction of its length,  $\langle\Delta R_\parallel^{\thinspace 2}\rangle/L$ approaches its limiting value from above, so that $P$, as given by Eq. (\ref{P2}), approaches its limiting value from below. In Fig. 2 the lower and upper curves in the inset show the finite size corrections for polymers with one free end and two free ends, respectively, with the latter case corresponding to the experiment. For $P=10$ $\mu$m, $D_x=4$ $\mu$m, $L=21$ $\mu$m, the rescaled length $t'=(2P)^{-1/3}D_x^{2/3}L$ is about 3.1, and for this value of $t'$, $\langle\Delta R_\parallel^{\thinspace 2}\rangle/L\propto w^2t'$ is seen to be about 50 $\%$ larger than its large $t'$ limit. The actual finite-size corrections are expected to be even larger than this, since Fig. 2 is based on the Hamiltonian (\ref{hamiltonian}), which is equivalent to the worm-like chain for small slopes $|d\vec{\bf r}/dt|\ll 1$ , but for larger slopes overestimates the bending energy \cite{explain}.

In comparing the estimates of  $P$ from Eqs. (\ref{P1}) and (\ref{P2}), one should keep in mind that the prediction of Eq. (\ref{P1}) is extremely sensitive to the experimental uncertainty in the normalized mean $\langle\Delta R_\parallel\rangle/L$, since this quantity is close to unity for a long tightly-confined polymer, so that the denominator in Eq. (\ref{P1}) nearly vanishes. For example, increasing $\langle\Delta R_\parallel\rangle/L$ from the value 0.93 by 3 \% {\em more than doubles} the estimate of $P$. In view of this, the disagreement of the numerical estimates based on Eqs. (\ref{P1}) and (\ref{P2}) mentioned a few paragraphs above is not so surprising. One advantage of Eq. (\ref{P2}) over Eq. (\ref{P1}) is that the relative uncertainties in $P$ and $\langle\Delta R_\parallel^{\thinspace 2}\rangle/L$ are the same.

\section{Concluding Remarks}
In Ref. \cite{ybg} and this paper we have determined the universal amplitudes $\alpha_\circ$, $\beta_\circ$, $\alpha_\Box$, and $\beta_\Box$ in the scaling forms (\ref{Rparallelcirc}), (\ref{DeltaRparallelcirc}), (\ref{RparallelBox}), and (\ref{DeltaRparallelBox}) for the worm-like chain in cylindrical channels with good precision from simulations. We hope the results will be useful in analyzing experiments. Combining measurements of $\langle R_\parallel\rangle$ and $\langle \Delta R_\parallel^{\thinspace 2}\rangle$ and our predictions, one obtains two independent predictions for the persistence length $P$, which can be checked for consistency. We recall that $\langle\Delta R_\parallel^{\thinspace 2}\rangle$ may be determined by measuring the isothermal extension of a polymer in a channel placed under a weak tension \cite{springconstant} as well as by direct observation of the endpoint fluctuations.

We have also derived exact analytic results for a polymer confined by a parabolic potential rather than a hard wall and shown that $\langle \Delta R_\parallel^{\thinspace 2}\rangle$ is overestimated by about 30 $\%$ if the potential parameters are chosen to reproduce $L-\langle R_\parallel\rangle$ for a channel with hard walls.

Finally, we have compared our predictions for the scaling regime with the experimental data of Ref. \cite{kp} for the radial distribution function. The comparison points to a persistence length smaller than the values 13 $\mu$m and $15\pm 3$ $\mu$m reported in Refs. \cite{kp} and \cite{lm}, respectively, but the experimental parameters are at the edge or outside the scaling regime, and significant corrections to scaling are expected. For a more conclusive comparison with our results, we would welcome experiments that probe deeper into the scaling regime $D_x,D_y\ll P\ll L$.

\appendix
\section{Calculational details for parabolic confining potential}

For the Hamiltonian (\ref{hamiltonian}) with the potential energy (\ref{parabolicpotential}), the polymer partition function $Z^{(3)}_L$ for a polymer in the three dimensional space $(x,y,t)$ factors in the form
\begin{equation}
Z^{(3)}_L(a_x,b_x;a_y,b_y)=Z^{(2)}_L(a_x,b_x)Z^{(2)}_L(a_y,b_y)\;.\label{Z}
\end{equation}
Here
\begin{equation}
Z^{(2)}_L(a,b)=\int Dx\exp\left\{-{1\over 2}\int_0^L dt\left[P\left({d^2x\over dt^2}\right)^2+a\thinspace  \left({dx\over dt}\right)^2+b\thinspace x^2\right]\right\}\label{Z2}
\end{equation}
is the partition function of a worm-like chain in two spatial dimensions $(x,t)$, with a parabolic confining potential.

In Eqs. (\ref{Z}) and (\ref{Z2}), auxiliary fields $a_x$ and $a_y$ have been introduced for conveniently generating correlations of $\int_0^L dt\thinspace v_x^2$ and $\int_0^L dt\thinspace v_y^2$ by differentiation. The auxiliary fields have a physical interpretation related to tension. If one end of the polymer is fixed and the other end is free to move but subject to a force or tension $\tau$ applied in the longitudinal direction, the corresponding potential energy $-\tau(R_\parallel-L)\approx {\tau\over 2}\int_0^L dt\;\vec{v}(t)^{\;2}$, where we have used Eq. (\ref{extension}), is included in the Hamiltonian and contributes to the Boltzmann factor. Comparing with the partition functions in Eqs. (\ref{Z}) and (\ref{Z2}), we see that $a_x=a_y=\tau/k_BT$.

For calculating ``bulk" properties of long polymers that are independent of the detailed boundary conditions at the ends, the periodic boundary condition $\vec{r}(t)=\vec{r}(t+L)$ is especially convenient. With the substitution $x(t)=L^{-1/2}\sum_q x_q e^{iqt}\;$, Eq. (\ref{Z2}) takes the form
\begin{equation}
Z^{(2)}_L(a,b)=\int Dx \prod_q\exp\left[-{1\over 2}\left(Pq^4+aq^2+b\right)x_q x_{-q}\right]\;.\label{Z22}
\end{equation}
The subtracted free energy $\Delta f^{(2)}(a,b)$, defined by
\begin{equation}
{\Delta f^{(2)}(a,b)\over k_BT}=-L^{-1}\ln\left[Z^{(2)}_L(a,b)/Z^{(2)}_L(0,0)\right]\;,\label{Deltaf}
\end{equation}
may be evaluated by standard Gaussian integration techniques \cite{joyce} and is given by
\begin{equation}
{\Delta f^{(2)}(a,b)\over k_BT}=\int_0^\infty{dq\over 2\pi}\thinspace\ln{Pq^4+aq^2+b\over Pq^4}=2^{-1/2}b^{1/4}P^{-1/4}\left(1+{a\over 2\sqrt{bP}}\right)^{1/2}\;.\label{Deltaf2}
\end{equation}

The right-most expression in Eq. (\ref{Deltaf2}) also follows readily from the path-integral approach of Ref. \cite{twb95}, according to which the partition function of the polymer with fixed endpoints and endslopes has the expansion
\begin{equation}
Z^{(2)}(x,v;x_0,v_0;L)=\sum_\nu\psi_\nu(x,v)\psi_\nu(x_0,-v_0)e^{-E_\nu\thinspace L}\;,\label{propagator}
\end{equation}
analogous to a quantum mechanical propagator. The eigenvalues and eigenfunctions are solutions of the $L$-independent Fokker-Planck equation
\begin{equation} \left(v{\partial\over\partial
x}-{1\over 2P}\thinspace{\partial^2\over
\partial v^2}+{1\over 2}\thinspace b x^2+{1\over 2}\thinspace a v^2\right)\psi(x,v)=E\thinspace \psi(x,v)\;.\label{fp}
\end{equation}
The dominant contribution for large $L$ in Eq. (\ref{propagator}) comes from the ground state, which has eigenfunction $\psi_0(x,v)$ and eigenvalue $E_0$, where $E_0=(k_BT)^{-1}\Delta f^{(2)}(a,b)$, as follows from Eqs. (\ref{Deltaf}) and (\ref{propagator}). According to Ref. \cite{twb95}, $\psi_0(x,v)$ has the Gaussian form $\psi_0(x,v)=A\exp\left(-Bx^2-Cxv-Dv^2\right)$. Requiring that this expression satisfy Eq. (\ref{fp}) determines $E_0$ and the constants $B$, $C$, $D$, and  Eq. (\ref{fp}), yielding
\begin{equation}
\psi_0(x,v)=A\exp\left[-(bP)^{1/2}E_0\thinspace x^2+(bP)^{1/2} xv-P E_0\thinspace v^2\right]\;,
\end{equation}
with $E_0$ given by the right-most expression in  Eq. (\ref{Deltaf2}).

Setting $a$=0 in Eq. (\ref{Deltaf2}) and including the  contributions from displacements in both the $x$ and $y$ directions into account, we obtain the free energy per unit length of confinement in Eq. (\ref{Deltafparabolic}), which is consistent with Eq. (16) of Ref. \cite{twb95}.

From Eqs. (\ref{Z2}), (\ref{Deltaf}), and (\ref{Deltaf2}),
\begin{equation}
{\partial\over\partial a}\thinspace{\Delta f^{(2)}(a,b)\over k_BT}={1\over 2L}\int_0^L dt\thinspace\langle v(t)^2\rangle=
2^{-5/2}b^{-1/4}P^{-3/4}\left(1+{a\over 2\sqrt{bP}}\right)^{-1/2}\;.\label{1stderiv}
\end{equation}
To calculate the average longitudinal extension $\langle R_\parallel\rangle$, we set $a$=0 in Eq. (\ref{1stderiv}), substitute the result in Eq. (\ref{Rparallelav}), and include the  contributions from transverse displacements in both the $x$ and $y$ directions. This yields the expression for the average longitudinal extension given in Eq. (\ref{Rparallelavparabolic}).

Similarly, from Eqs. (\ref{Z2}), (\ref{Deltaf}), and (\ref{Deltaf2}),
\begin{eqnarray}
&&-\left({\partial\over\partial a}\right)^2{\Delta f^{(2)}(a,b)\over k_BT}={1\over 4L}\int_0^L dt_1\int_0^L dt_2\thinspace\left[\langle v(t_1)^2 v(t_2)^2\rangle-\langle v(t_1)^2\rangle\langle v(t_2)^2\rangle\right]\nonumber\\
&&\qquad =2^{-9/2}b^{-3/4}P^{-5/4}\left(1+{a\over 2\sqrt{bP}}\right)^{-3/2}\;.\label{2ndderiv}
\end{eqnarray}
To obtain $\langle \Delta R_\parallel^{\thinspace 2}\thinspace\rangle$ , we set $a$=0 in Eq. (\ref{2ndderiv}), substitute the result in Eq. (\ref{meansquare}), and include the  contributions from tranverse displacements in both the $x$ and $y$ directions. This yields the expression for the average longitudinal extension given in Eq. (\ref{meansquareparabolic}).

It is straightforward to derive the complete distribution function
\begin{equation}
P^{(2)}(\xi;a,b)=\left\langle\delta\left(\xi-{1\over 2}\int_0^Ldt\thinspace v(t)^2\right)\right\rangle\;,\label{P(z)}
\end{equation}
from which the above moments follow. Its Laplace transform is given by
\begin{equation}
\tilde{P}^{(2)}(s;a,b)=\int_0^\infty d\xi\thinspace e^{-s\xi}P(\xi;a,b)=\left\langle\exp\left(-{s\over 2}
\int_0^Ldt\thinspace v(t)^2\right)\right\rangle\;,\label{Ptilde1}
\end{equation}
where the average is to be carried out with the same Boltzmann weight as in Eq. (\ref{Z2}).
Thus,
\begin{equation}
\tilde{P}^{(2)}(s;a,b)={Z_L^{(2)}(a+s,b)\over Z_L^{(2)}(a,b)}=\exp\left[-L\thinspace{f^{(2)}(a+s,b)-f^{(2)}(a,b)\over k_BT}\right],
\label{Ptilde2}
\end{equation}
where we have made use of the definition (\ref{Deltaf}). Substituting Eq. (\ref{Deltaf2}) in Eq. (\ref{Ptilde2}) and evaluating the inverse Laplace transform, we find that $P^{(2)}(\xi;a,b)$ is given by the ``inverse Gaussian"
or Wald distribution \cite{invgaussdist}
\begin{equation}
P_{\rm invgauss}(\xi)={1\over \sqrt{2\pi\langle\Delta\xi^2\rangle}}\thinspace\left({\langle\xi\rangle\over\thinspace\xi}\right)^{3/2}
\exp\left[-{\langle\xi\rangle\over\thinspace\xi}\thinspace{(\xi-\langle\xi\rangle)^2\over 2\langle\Delta\xi^2\rangle} \right]\;,\quad 0<\xi <\infty\;,\label{invgauss}
\end{equation}
where
\begin{eqnarray}
&&\langle\xi\rangle_{a,b}=2^{-5/2}b^{-1/4}P^{-3/4}\left(1+{a\over 2\sqrt{bP}}\right)^{-1/2}L\;,\label{mean}\\
&&\langle\Delta\xi^2\rangle_{a,b}=2^{-9/2}b^{-3/4}P^{-5/4}\left(1+{a\over 2\sqrt{bP}}\right)^{-3/2}L\;,\label{variance}
\end{eqnarray}
are the mean and variance of the distribution, respectively, consistent with Eqs. (\ref{1stderiv}) and (\ref{2ndderiv}). Note that inverse Gaussian distribution vanishes as $\xi$ approaches zero, as expected from Eq. (\ref{P(z)}), reflecting the fact that the end-to-end distance $R_\parallel$ of the polymer cannot exceed the contour length.

Since the mean and variance in Eqs. (\ref{mean}) and (\ref{variance}) are both proportional to $L$, the inverse Gaussian distribution (\ref{invgauss}) reduces to the ordinary Gaussian form
\begin{equation}
P_{\rm gauss}(\xi)={1\over \sqrt{2\pi\langle\Delta\xi^2\rangle}}\thinspace
\exp\left[-{(\xi-\langle\xi\rangle)^2\over 2\langle\Delta\xi^2\rangle} \right]\;,\quad -\infty<\xi <\infty\;,\label{gausslim}
\end{equation}
in the large-$L$ limit.

The above results for a polymer in a two dimensional space $(x,t)$ are easily generalized to three spatial dimensions.
In Eq. (\ref{P(z)}), the quantity $v^2$ is replaced by $v_x^2+v_y^2$, so that
\begin{equation}
\xi=L-R_\parallel\;,\label{xi3d}
\end{equation}
in agreement with Eq. (\ref{extension}), and Eq. (\ref{Ptilde2}) is replaced by
\begin{equation}
\tilde{P}^{(3)}(s;a_x,b_x;a_y,b_y)={Z_L^{(2)}(a_x+s,b_x)\over Z_L^{(2)}(a_x,b_x)}\thinspace{Z_L^{(2)}(a_y+s,b_y)\over Z_L^{(2)}(a_y,b_y)}\;.\label{Ptilde3}
\end{equation}
Accordingly, the inverse Laplace transform is given by the convolution
\begin{equation}
P_\Box^{(3)}(\xi)=\int_0^{\xi}d\xi'\thinspace P^{(2)}(\xi-\xi';a_x,b_x)P^{(2)}(\xi';a_y,b_y)\;,\label{PBox(xi)}
\end{equation}
where each of the factors $P^{(2)}$ in the integrand has the inverse Gaussian form (\ref{invgauss}), with mean and variance defined by Eqs. (\ref{mean}) and (\ref{variance}).

In the case of cylindrically symmetric potential parameters $a_x=a_y=a$, $b_x=b_y=b$, appropriate for a channel with a circular or square cross section, the convolution in Eq. (\ref{PBox(xi)}) can be evaluated (or circumvented). The corresponding distribution also has the inverse Gaussian form \begin{equation}
P_\circ^{(3)}(\xi)=P_{\rm invgauss}(\xi)\;,\quad\langle\Delta\xi\rangle=2\langle\Delta\xi^2\rangle_{a,b}\;,\quad \langle\Delta\xi^2\rangle=2\langle\Delta\xi^2\rangle_{a,b}\;,\quad 0<\xi <\infty\;,\label{Pcirc(xi)}
\end{equation}
in terms of the distribution (\ref{invgauss}) and the mean and variance defined in Eqs. (\ref{mean}) and (\ref{variance}).

In the large-$L$ limit, in which $P^{(2)}(\xi;a,b)$ becomes Gaussian, the distribution functions $P^{(3)}_\Box(\xi)$ and $P^{(3)}_\circ(\xi)$ both take the Gaussian form (\ref{gausslim}), with mean $\langle\xi\rangle=\langle\xi\rangle_{a_x,b_x}+\langle\xi\rangle_{a_y,b_y}$ and variance $\langle\Delta\xi^2\rangle=\langle\Delta\xi^2\rangle_{a_x,b_x}+\langle\Delta\xi^2\rangle_{a_y,b_y}$ defined in Eqs. (\ref{mean}) and (\ref{variance}), as is consistent with Eqs. (\ref{Rparallelavparabolic}) and (\ref{meansquareparabolic}).

Like Eqs. (\ref{Deltaf}) and (\ref{Ptilde2}), our predictions (\ref{PBox(xi)}) and (\ref{Pcirc(xi)}) for the distributions $P_\Box(\xi)$ and $P_\circ(\xi)$ in terms of inverse Gaussian functions are really only exact in the large-$L$ limit, in which the ground-state contribution to the sum in Eq. (\ref{propagator}) dominates. However, for moderately large $L$ the distributions also work quite well, reproducing the skewed form of the radial distribution observed experimentally and calculated theoretically in Refs. \cite{lm,twf}. This is shown in Section IV, where our results are compared with recent experimental data of K\"oster and Pfohl \cite{kp} for the radial distribution function.

Finally we argue that the distribution of $R_\parallel$ becomes Gaussian in the large-$L$ limit not just for the parabolic potential, but for general confining potentials, including the hard-wall potential. To see this, note that for a general confining potential, the Laplace transform of the distribution function, defined as in Eqs. (\ref{P(z)}) and (\ref{Ptilde1}) is related to the free energy per unit length $f(a)$ by
\begin{eqnarray}
\tilde{P}^{(2)}(s;a)&=&\exp\left[-L\thinspace{f^{(2)}(a+s)-f^{(2)}(a)\over k_BT}\right]\nonumber\\
&=&\exp\left[-\langle\xi\rangle\thinspace s+{1\over 2}\langle\Delta\xi^2\rangle\thinspace s^2+L\thinspace{\rm O}(s^3)\right]\;,\label{Ptildegen}
\end{eqnarray}
analogous to Eq. (\ref{Ptilde2}). Here we have expanded $f(s+a)$ to second order in $a$, relating the expansion coefficients to moments of $\xi$, as above. With the substitution $s=iy$ the inverse Laplace transform of Eq. (\ref{Ptilde2}) takes the form
\begin{equation}
\tilde{P}^{(2)}(\xi;a)={1\over 2\pi}\int_{-\infty}^\infty dy\thinspace\exp\left[i(\xi-\langle\xi\rangle)\thinspace y
-{1\over 2}\langle\Delta\xi^2\rangle\thinspace y^2+L\thinspace{\rm O}\left((iy)^3\right)\right]\;.\label{Pinvtransform}
\end{equation}
Treating the ${\rm O}\left((iy)^3\right)$ term in square brackets perturbatively, one finds a negligible contribution, for large $L$, to the Gaussian distribution (\ref{gausslim}) implied by the first two terms.

\acknowledgments TWB thanks Theo Odijk for correspondence, Thomas Pfohl for sending the experimental data considered in Section IV,  Dieter Forster for a helpful discussion, and Robert Intemann for help with {\it Mathematica}. YY acknowledges financial support from the International Helmholtz Research School``BioSoft".

\newpage
\begin{figure}[Figure1]
\begin{center}
%\begin{minipage}{\textwidth}
%\includegraphics[width=18cm]{probdist-xix.eps}
\includegraphics[width=6in]{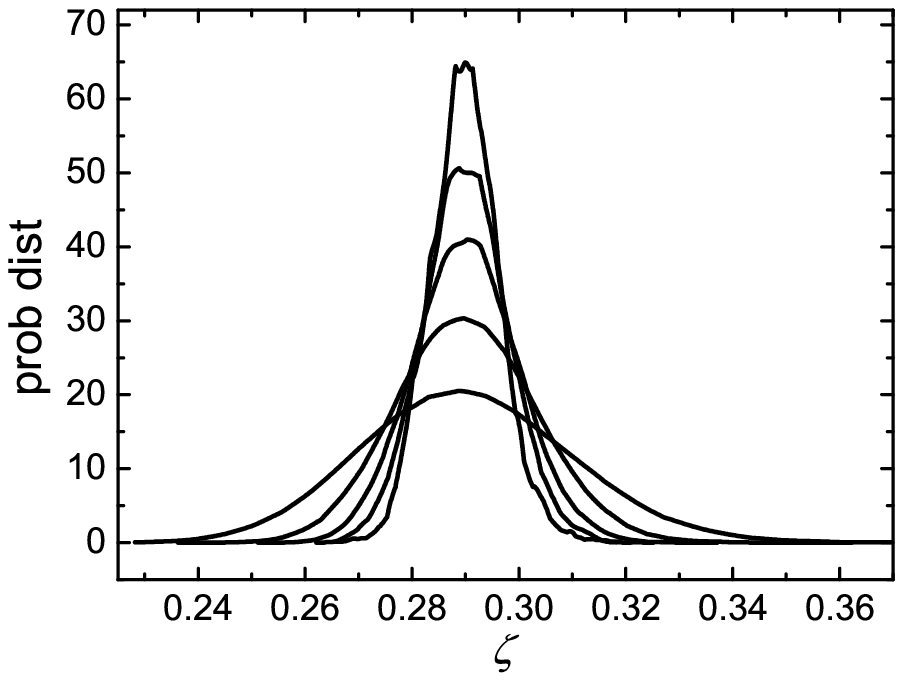}
%\end{minipage} \\
\caption{Distribution of the quantity $\zeta=t'^{-1}\int_0^{t'}dt'\thinspace v_x'^2$ for a rescaled polymer with persistence length $P'={1\over 2}$ and longitudinal length $t'$ on a two-dimensional strip of width 1. The curves correspond, from bottom to top, to $t'=$ 100, 225, 400, 625, and 900.}\label{fig1}
\end{center}
\end{figure}

\newpage
\begin{figure}[Figure2]
\begin{center}
%\begin{minipage}{\textwidth}
\includegraphics[width=6in]{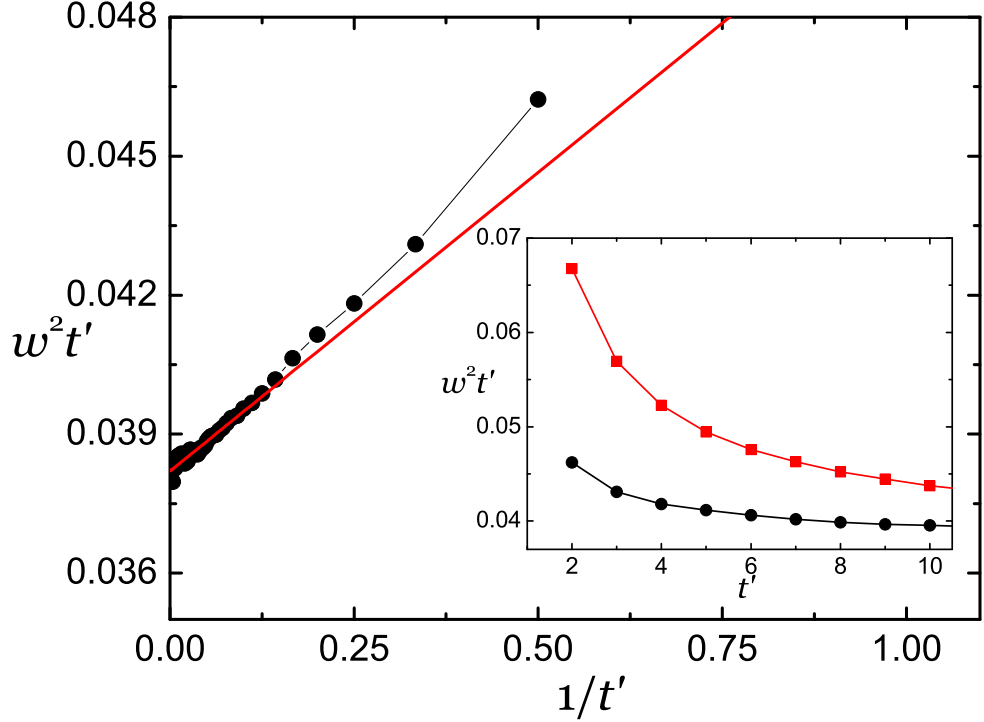}
%\end{minipage} \\
\caption{Dependence of $w^2 t'$ on $t'^{-1}$, where $w^2$ is the variance or mean square deviation of the distribution in Fig. 1. The straight line shows the best two-parameter fit of the data from $t'=11$ to 300 to the functional form $w^2 t'=k+\ell\thinspace t'^{-1}$, for which $k=0.0382$ and $\ell=0.0129$. The round points are simulation results for a polymer with one end fixed in the middle of the strip with slope $\vec{v_0}=0$ and with the other end free to fluctuate. The square points are results for a polymer with both ends free to fluctuate, as in the experiments. As seen in the inset, the finite size corrections to the limiting value for large $t'$ are greater in the case of two free ends.}\label{fig2}
\end{center}
\end{figure}

\newpage
\begin{figure}[Figure3]
\begin{center}$
\begin{array}{cc}
\includegraphics[width=3.5in]{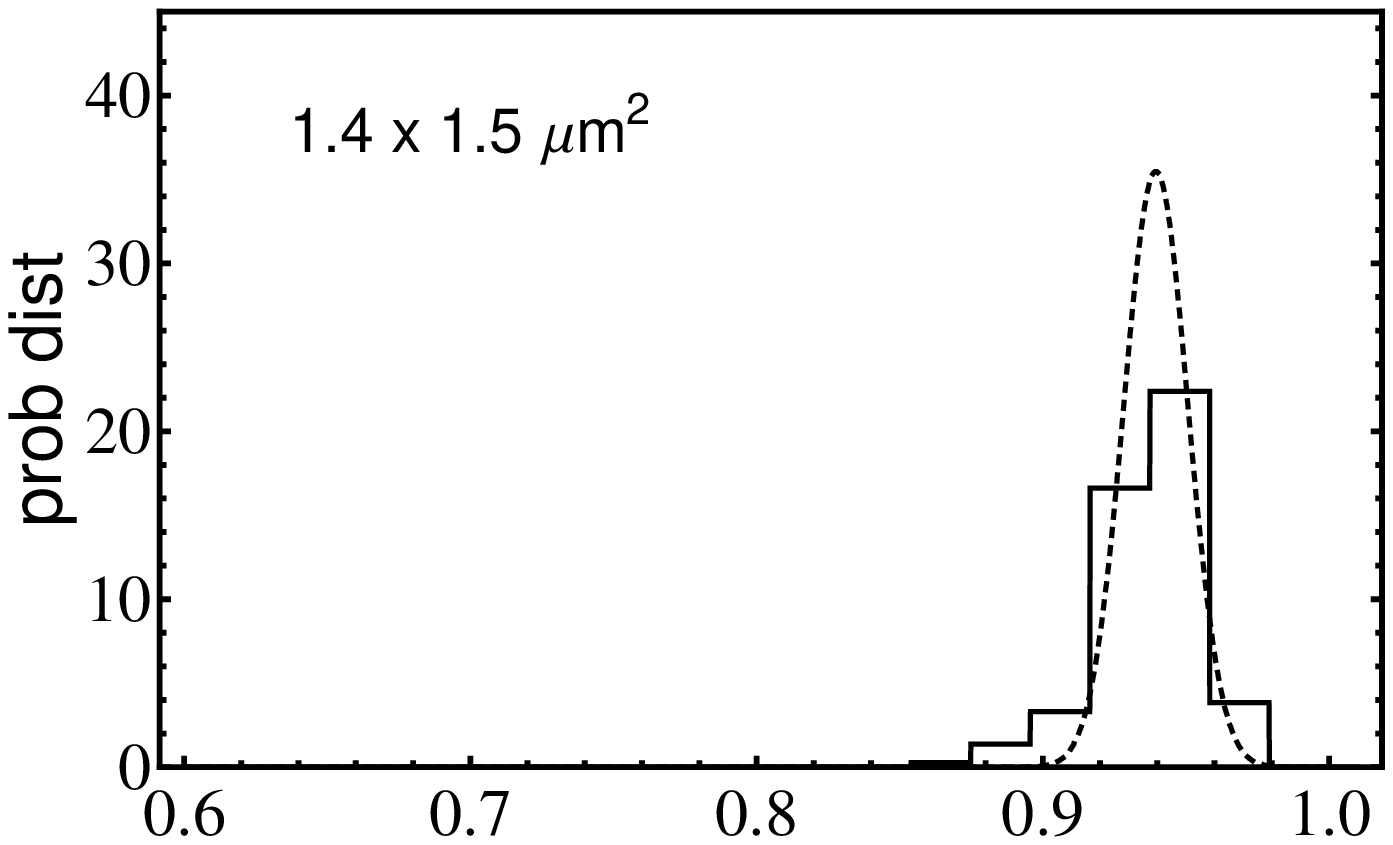}&
\includegraphics[width=3.5in]{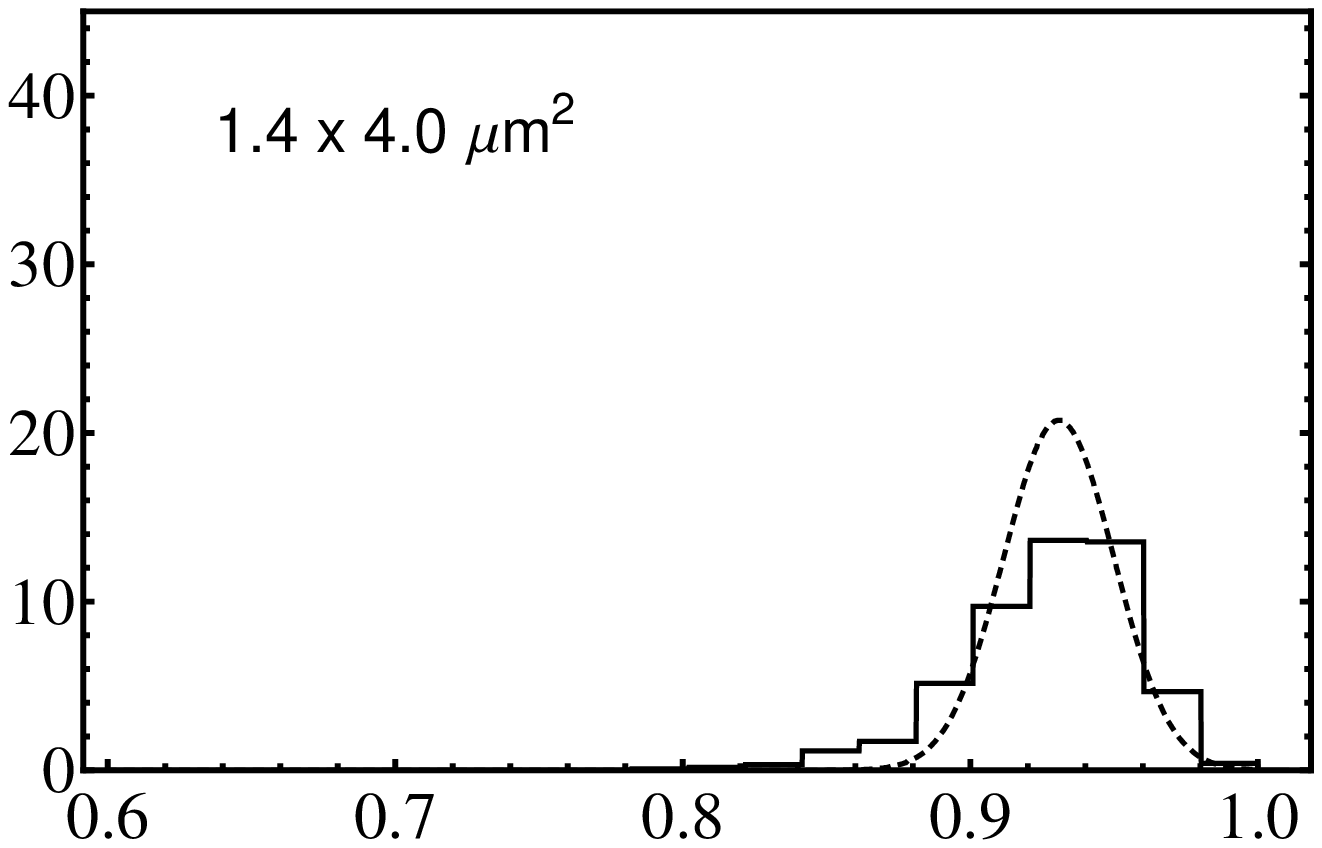}\\
\includegraphics[width=3.5in]{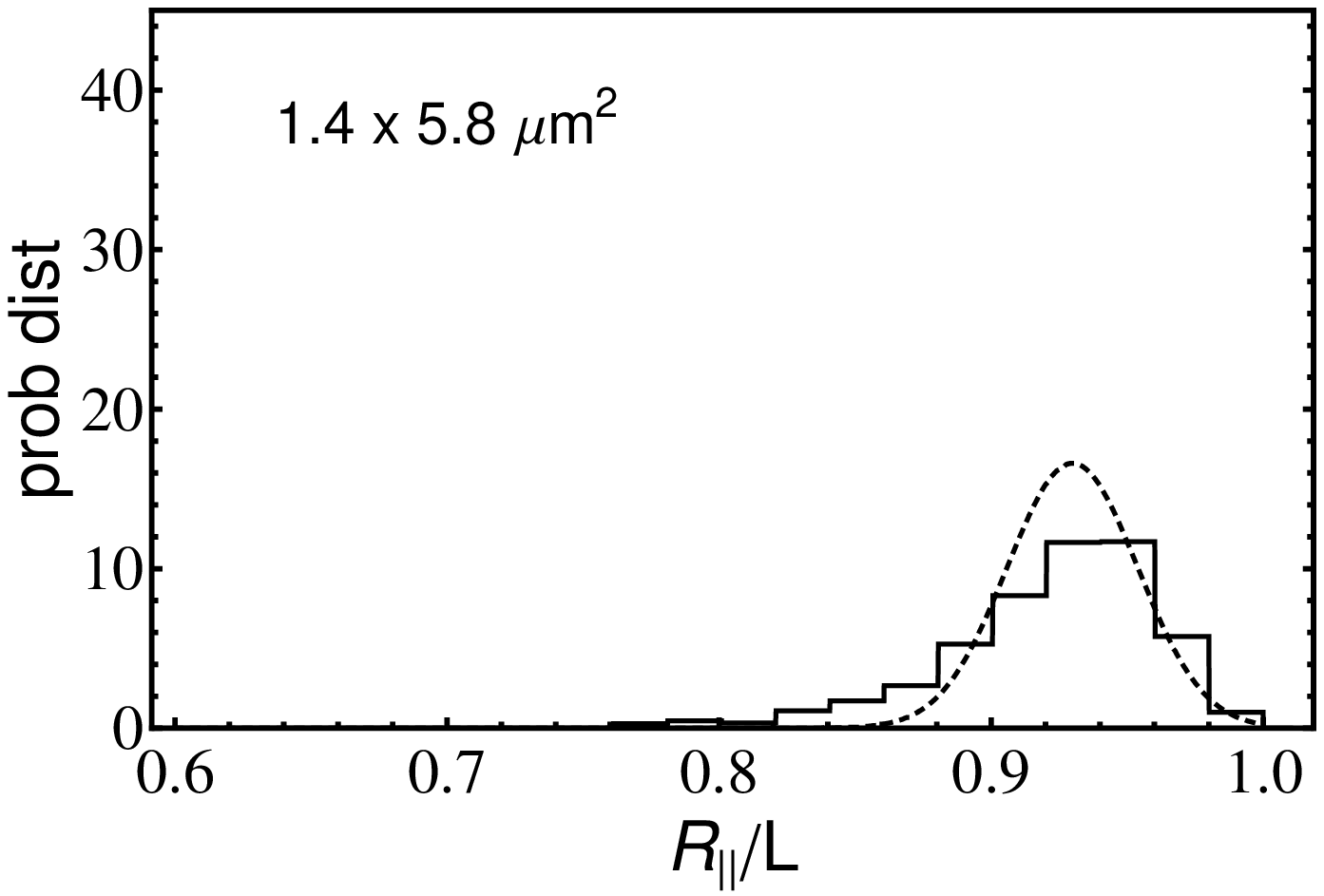}&
\includegraphics[width=3.5in]{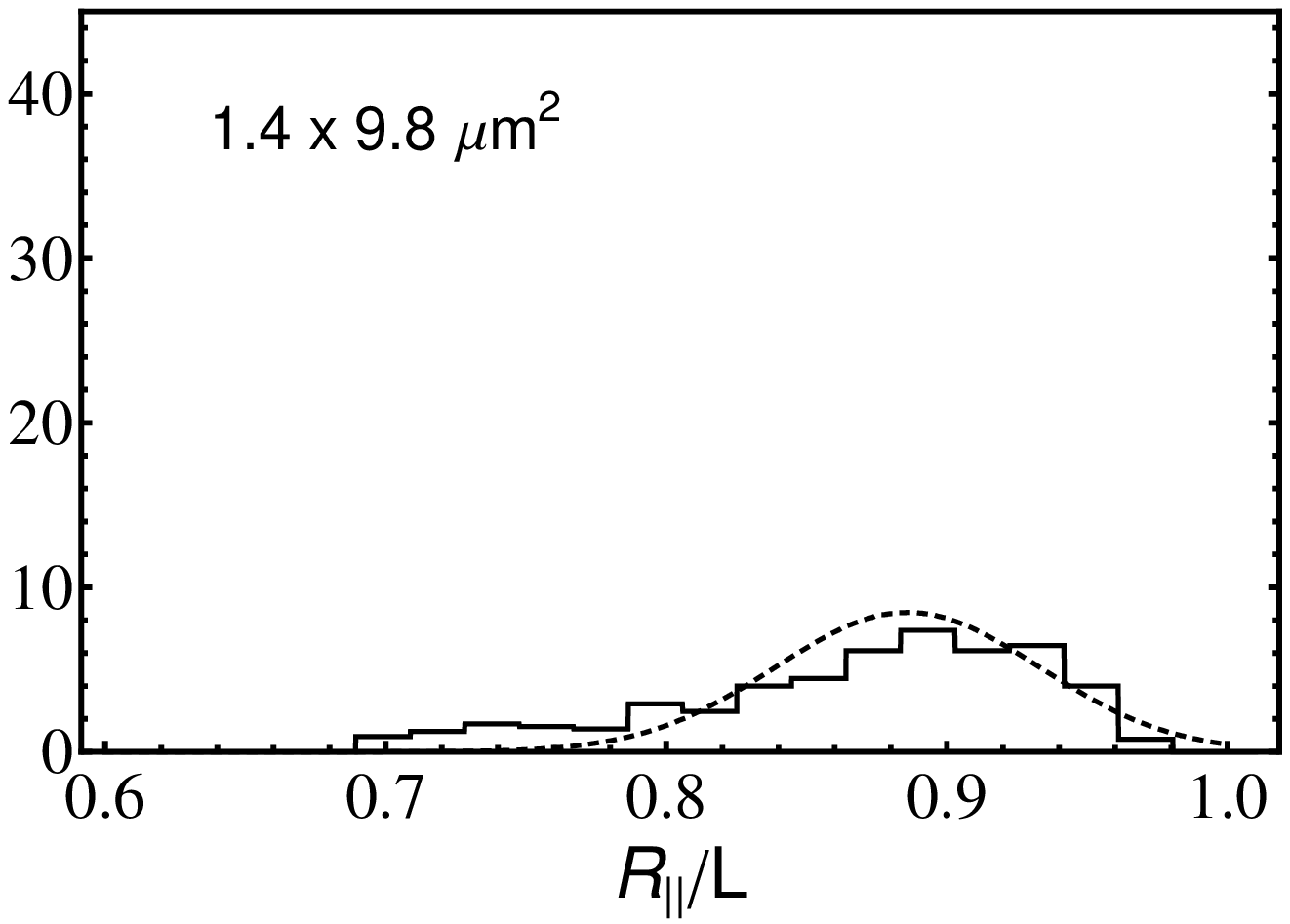}
\end{array}$
\caption{One-parameter least squares fit of the experimental results (histogram) of K\"oster and Pfohl \cite{kp} for the radial distribution function of actin filaments in channels with cross section $D_x\times D_y$ with the expected Gaussian distribution (dashed curves) for $D_x,D_y\ll P\ll L$. The histograms and dotted curves are normalized to unit area. The mean and variance of the Gaussian curves were determined from Eqs. (\ref{RparallelBox}) and (\ref{DeltaRparallelBox}), using the estimates of $\alpha_\Box$ and $\beta_\Box$ in Eqs. (\ref{alphaBoxcirc}) and (\ref{betaBoxcirc}), for the experimental parameters $L=21\;\mu$m, $D_x=1.4\;\mu$m, and $D_y=$1.5, 4.0, 5.8, and $9.8\;\mu$m. Choosing the persistence length $P$ to optimize the fit, yields the estimates $P=$ 7.61, 11.1, 14.1, and 10.1 $\mu$m for the channels of width 1.5, 4.0, 5.8, and 9.8 $\mu$m.} \label{fig3}
\end{center}
\end{figure}

\newpage
\begin{figure}[Figure4]
\begin{center}$
\begin{array}{cc}
\includegraphics[width=3.5in]{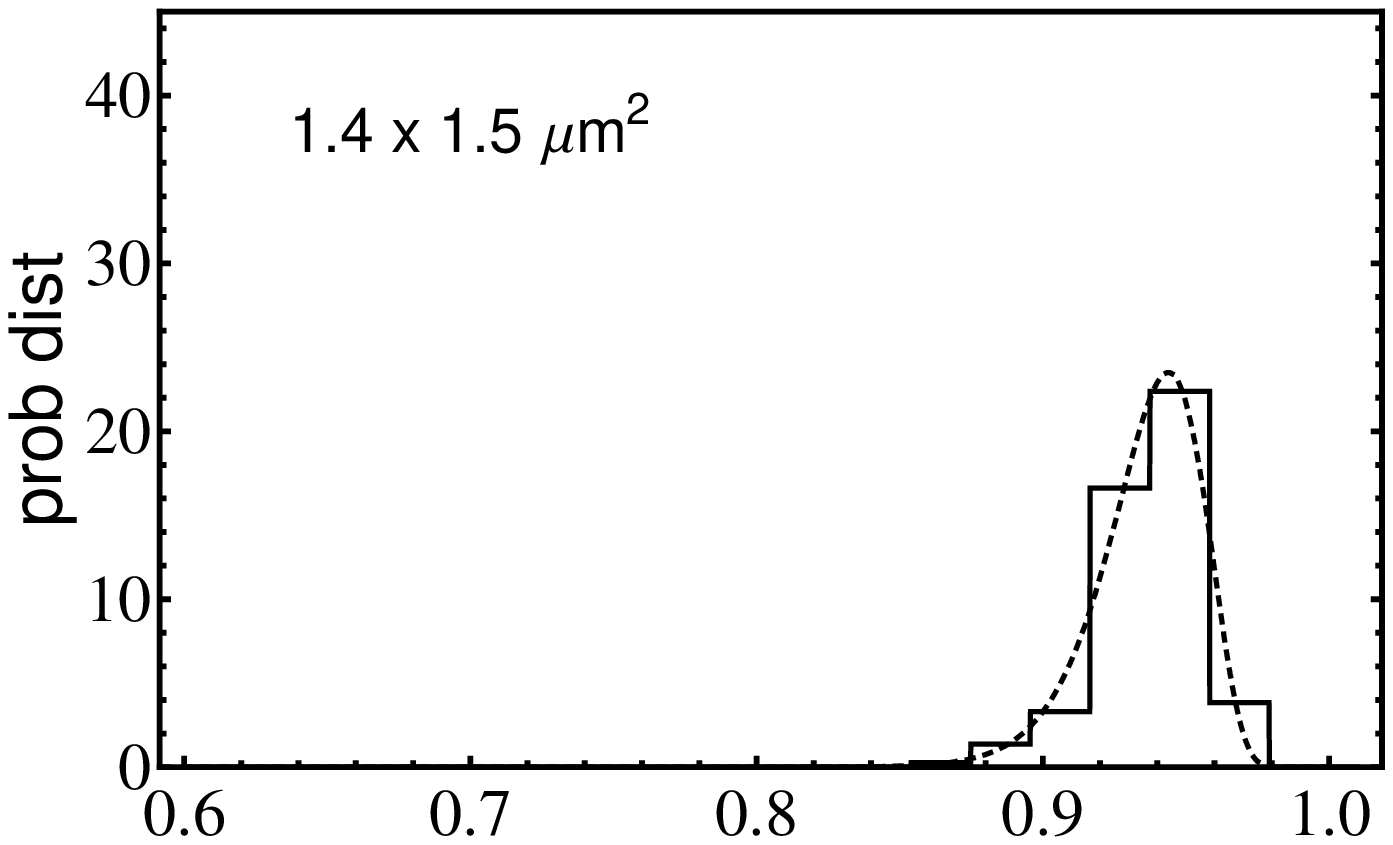}&
\includegraphics[width=3.5in]{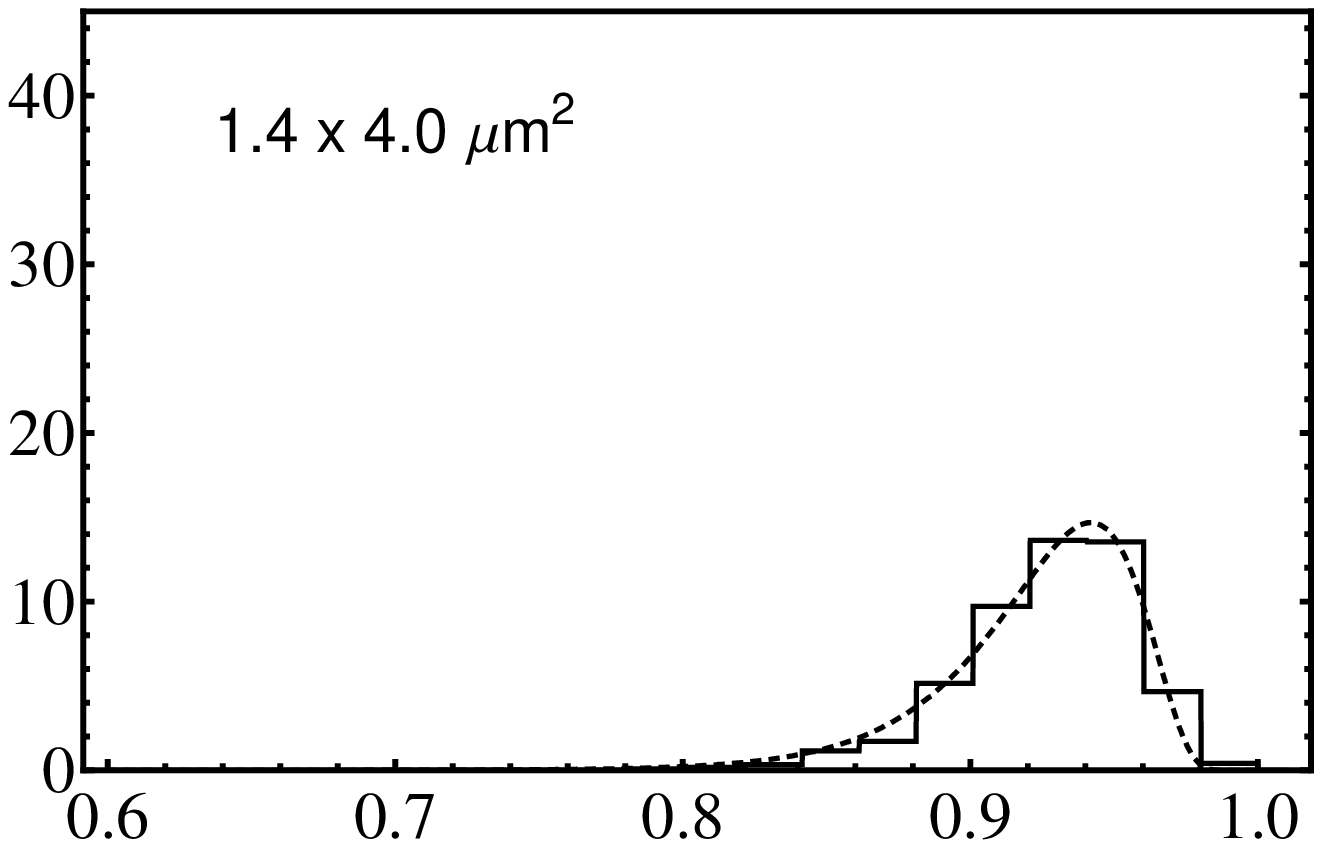}\\
\includegraphics[width=3.5in]{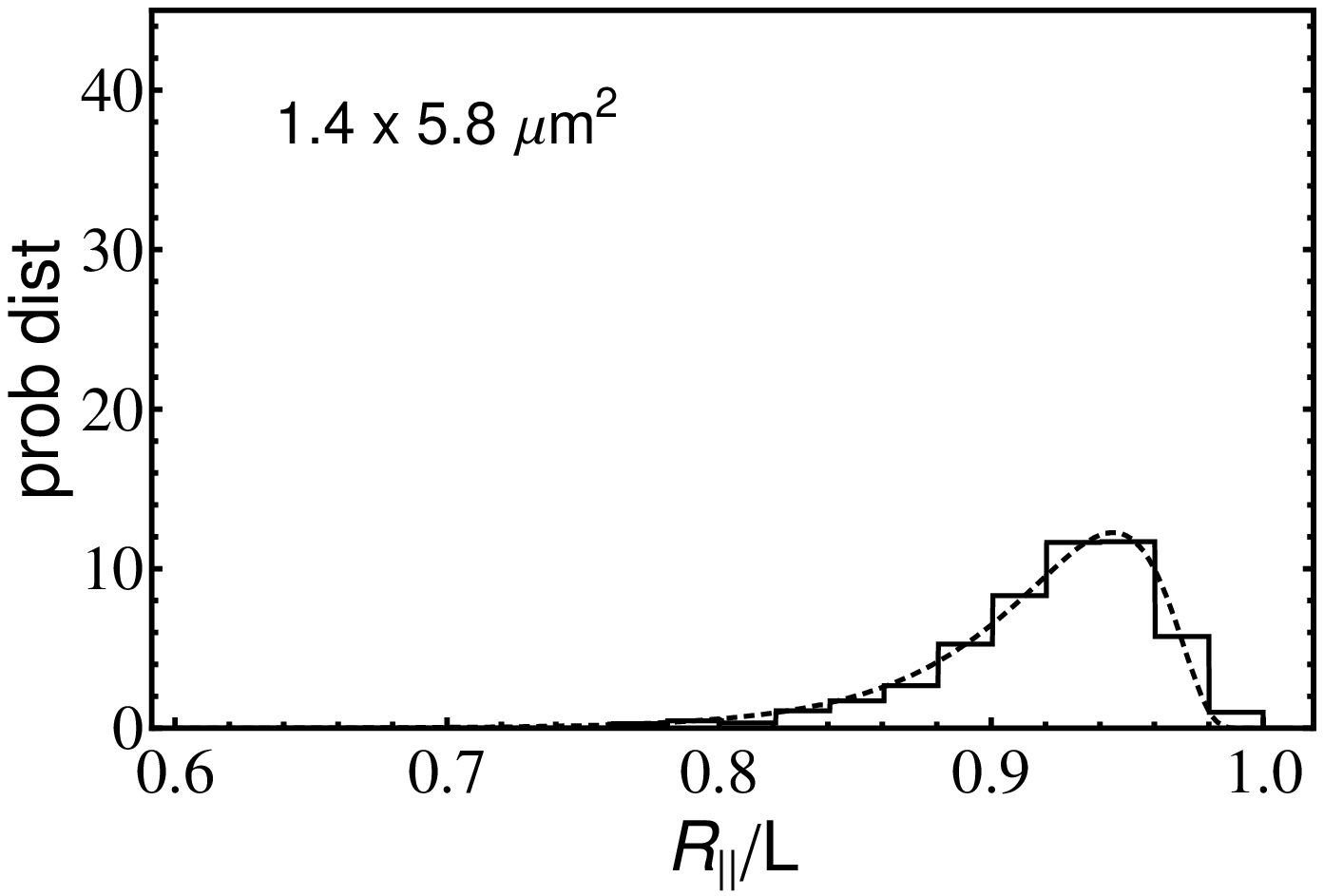}&
\includegraphics[width=3.5in]{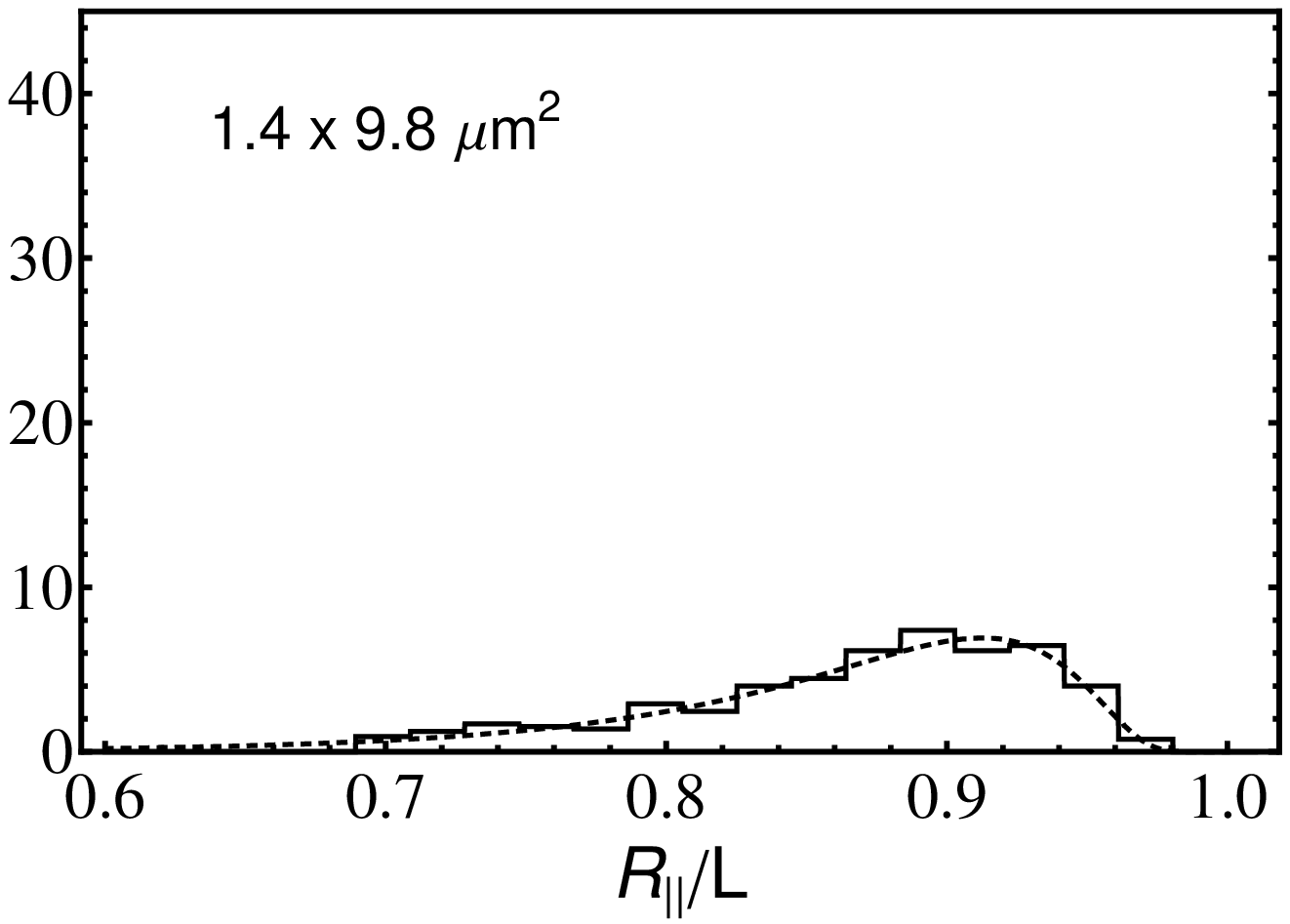}
\end{array}$
\caption{Two parameter fit of the same experimental data for the radial distribution function as in Fig. 3 to the inverse Gaussian distribution, given by Eqs. (\ref{invgauss}) and (\ref{xi3d}), with both the mean $\langle R_\parallel\rangle$ and variance $\langle\Delta R_\parallel^{\thinspace 2}\rangle$ chosen to optimize the fit. This leads to the estimates $P=(7.0,\thinspace 2.8)$, $(9.5,\thinspace 3.6)$, $(10.8,\thinspace 4.1)$, and $(7.2,\thinspace 3.2)$ in $\mu$m for the channel widths $D_y=1.5$, 4.0, 5.8, and 9.8 $\mu$m, where the first and second numbers in parenthesis follow from Eqs. (\ref{P1}) and (\ref{P2}), respectively, with $\alpha_\Box$ and $\beta_\Box$ given by Eqs. (\ref{alphaBoxcirc}) and (\ref{betaBoxcirc}).}\label{fig4}
\end{center}
\end{figure}

\begin{thebibliography}{99}
\bibitem{explain} The bending energy ${\cal H}_{\rm bend}$ of a worm-like chain  is
${\cal H}_{\rm bend}/k_BT={1\over 2}P\int_0^L ds\thinspace\left(d\hat{\tau}/ds\right)^2$,
where $\hat{\tau}=d\vec{r}/ds$ is a unit tangent vector and $s$ denotes the arc length.
Rewritten in terms of the quantities $\vec{v}=d\vec{r}/dt$ and $\vec{a}=d^2\vec{r}/dt^2$, ${\cal H}_{\rm bend}/k_BT={1\over 2}P\int_0^L dt
\left[(1+\vec{v}^2)^{-3/2}\thinspace\vec{a}^2-(1+\vec{v}^2)^{-5/2}\thinspace(\vec{v}\cdot\vec{a})^2\right]$. For $|\vec{v}|\ll 1$, this leads to Eq. (\ref{hamiltonian}).
\bibitem{explainwidth}
Comparing Eqs. (\ref{meansquare}) and (\ref{w}), one sees that $w^2$ varies like $L^{-2}\langle\Delta R_\parallel^{\thinspace 2}\rangle$ for large $L$. According to the scaling forms (\ref{DeltaRparallelcirc}) and (\ref{DeltaRparallelBox}), $\langle\Delta R_\parallel^{\thinspace 2}\rangle$ is asymptotically proportional to $L$. Thus, $w\sim L^{-1}\sim t'^{\thinspace -1}$. That $\langle\Delta R_\parallel^{\thinspace 2}\rangle$ is proportional to $L$ also follows (i) from the property that the two point correlation function in Eq. (\ref{meansquare}) is short ranged and (ii) from the general arguments in the last paragraph of the Appendix.
\bibitem{springconstant} The effective spring constant is given by
$k_{\rm eff}= k_BT/\langle\Delta R_\parallel^{\thinspace 2}\rangle$. This rather general result, which holds for flexible and semiflexible polymers with and without confining geometries, follows from including the potential energy $-\tau R_\parallel$, where $\tau$ is the tension, in the Boltzmann weight and noting that $k_{\rm eff}^{-1}=\partial\langle R_\parallel\rangle/\partial\tau=(k_BT)^{-1}\langle\Delta R_\parallel^{\thinspace 2}\rangle$.
\end{thebibliography}
\end{document}